\documentclass[a4paper]{article}
\date{30.04.2019}

\usepackage{graphicx}
\usepackage[pdftex]{hyperref}
\usepackage{color}
\usepackage{amssymb}

\begin{document}


\title{Structural Properties of Irreducible Two-Particle Representations of the Poincar\'e Group}
\author{\bf{Walter Smilga}\footnote{Geretsried, Germany; wsmilga@compuserve.com}}


\maketitle

\begin{abstract}
Two particles, described by an irreducible two-particle representation of the Poincar\'e group, 
are correlated by the constraints that the constancy of the Casimir operators imposes on the state space.
This correlation can be understood as a geometrically caused interaction between the particles, the strength of 
which is related to the normalisation constant $\omega$ of the two-particle states by $4\pi\,\omega^2$.
The numerical value of $4\pi\,\omega^2$ is found to match the experimental value of the electromagnetic 
fine structure constant $\alpha$.
This strongly suggests that the correlation of two particles in an irreducible two-particle representation of 
the Poincar\'e group manifests itself in the electromagnetic interaction.
\\ \\
\noindent
\bfseries{Keywords}\mdseries:  
Poincar\'e group $\cdot$
Two-particle representation $\cdot$ Momentum entanglement $\cdot$ Fine-structure constant $\cdot$ Electromagnetic interaction $\cdot$ Geometrisation  
\end{abstract}

\section{Introduction} 
\label{INTRO}

There is a well-established connection between elementary particle physics and representation theory, as first 
noted in the 1930s by Wigner \cite{epw}. 
It links the properties of particles to the structure of Lie groups and Lie algebras.
An outstanding example is the Poincar\'e group, whose irreducible unitary representations provide the quantum 
mechanical state spaces of the various elementary particles.

According to the axioms of quantum mechanics, a system composed of independent particles is described by the 
tensor product of their individual state spaces, which means by a product representation of the Poincar\'e group.
In fact, the Standard Model (of particle physics) is based on product representations in connection with a Fock space 
formulation, which allows describing a variable number of particles. 
On the basis of these product representations, the electromagnetic, weak, and strong interactions are modelled by 
``interaction terms'' that couple the particles to gauge fields and thus establish correlations between the particles.

Another way of establishing correlations is to reduce a product representation to a direct sum of irreducible 
multi-particle representations.
An irreducible representation of the Poincar\'e group is characterised and labelled by fixed values of two 
Casimir operators. 
It describes an isolated multi-particle system, characterised by well-defined and conserved total linear and 
angular momenta. 
The particles of a multi-particle system described by such a representation are no longer independent, but are 
correlated by the constancy of the Casimir operators.

Both types of correlation, those by coupling to the gauge fields and those by fixing the values of the Casimir 
operators, can be described as an interaction mediated by the virtual exchange of momentum between the particles.
I will show that irreducible two-particle representations of the Poincar\'e group describe an interaction that is 
equal in structure and strength to the electromagnetic interaction.

\section{Irreducible unitary representations}
\label{IRRE}

\subsection{Basic properties}

Following the presentation in \cite{sss1}, I will briefly recall the basic properties of the irreducible unitary 
representations of the Poincar\'e group. (The speed of light $c$ and Planck's constant $\hbar$ are set to 1.) 
 
The Poincar\'e group is defined by the commutation relations of the infinitesimal generators of translations and Lorentz 
transformations, $p_\mu$ and $M_{\mu\nu}$, 
\begin{eqnarray}
\left[M_{\mu\nu}, p_\sigma \right] &=& i\,(g_{\nu\sigma} p_\mu - g_{\mu\sigma} p_\nu),       		\label{1-1} \\
\left[p_\mu, p_\nu \right] &=& 0,    																														\label{1-2} \\
\left[M_{\mu\nu}, M_{\rho\sigma}\right] &=& 
-i\,(g_{\mu\rho} M_{\nu\sigma} - g_{\nu\rho} M_{\mu\sigma} + 
g_{\mu\sigma} M_{\rho\nu} - g_{\nu\sigma} M_{\rho\mu}),  																				\label{1-3} 									
\end{eqnarray} 
where $g_{\mu\nu}$ is the Minkowski metric tensor $g = (g_{\mu\nu}) = $ diag$(1,-1,-1,-1)$.

In quantum mechanical applications, the infinitesimal generators $p_\mu$ and $M_{\mu\nu}$ are represented by self-adjoint 
linear operators, acting on a Hilbert space, which is the state space of the representation. 
These operators are identified with the following observables: their eigenvalues are real numbers that correspond 
to the measured values of linear momentum and angular momentum. 
(I will use the same notations for these operators as for the corresponding infinitesimal generators.)

The Poincar\'e group has two Casimir operators 
\begin{equation}
P = p^\mu p_\mu\;\; \mbox{and} \;\;																													
W = -w^\mu w_\mu\; , \; \mbox{ with } \;
w_\sigma = \frac{1}{2} \epsilon_{\sigma \mu \nu \lambda} M^{\mu \nu}p^\lambda; 									\label{1-4}
\end{equation}
here $\epsilon_{\sigma \mu \nu \lambda}$ is the antisymmetric tensor of rank 4 and
$w_\sigma$ is the Pauli--Lubanski pseudovector.
The commutation relations of $w_\sigma$ can be derived from the commutation relations (\ref{1-1}--\ref{1-3}):
\begin{eqnarray}
\left[M_{\mu\nu}, w_\rho \right] &=& i\,(g_{\nu\rho} w_\mu - g_{\mu\rho} w_\nu) ,         	  	\label{1-10}  \\
\left[w_\mu, p_\nu \right] &=& 0 ,  																														\label{1-11}	\\
\left[w_\mu, w_\nu \right] &=& i \epsilon_{\mu \nu \rho \sigma} w^\rho p^\sigma .								\label{1-12}
\end{eqnarray} 
With (\ref{1-10}--\ref{1-12}), it can then be verified that the Casimir operators (\ref{1-4}) commute with all the
infinitesimal generators $p_\mu$ and $M_{\mu\nu}$.

A representation is irreducible if the Hilbert space has no nontrivial invariant subspaces.  
In this case, the Casimir operators are multiples of the identity and their scalar values can be used to classify 
the irreducible representations. 

The eigenvalues of $P$ correspond to a squared mass $m^2$.
In a single-particle representation, relation $P = p^\mu p_\mu = m^2$ defines the ``mass shell'' of the particle.
In the following, I will only consider representations with $m^2 > 0$.

In vector notation, the Casimir operator $P$ can be written as
\begin{equation}
P = p_0^2 - \mathbf{p}^2,																																				\label{1-7}
\end{equation}			
where the vector $\mathbf{p}$ denotes the three spatial components of $p_\mu$.

The interpretation of the eigenvalues of the Casimir operator $W$ is less obvious.
The pseudovector $w_\sigma$ can be written as
\begin{equation}
w_0 = \mathbf{p} \cdot \mathbf{M}, \;\; \mathbf{w} = p_0\mathbf{M} - \mathbf{p} \times \mathbf{N}, 	\label{1-5}
\end{equation} 
where $\mathbf{M}=(M_{32}, M_{13}, M_{21})$ denotes the three components of the angular momentum and 
$\mathbf{N}=(M_{01}, M_{02}, M_{03})$ are the three boost operators.
It is convenient to make a Lorentz transformation to the rest frame in which $\mathbf{p}=\mathbf{0}$, $p_0=m$. 
Then (\ref{1-5}) reduces to
\begin{equation}
w_\mu = m\,(0, M_{23}, M_{31}, M_{12}) 		  						                        								\label{1-6}
\end{equation} 
and the Casimir operator $W$ takes on the simple form
\begin{equation}
W = m^2\,\mathbf{M}^2.																																					\label{1-8}
\end{equation}
The eigenvalues of $\mathbf{M}^2$ are $s(s+1)$, where $s = 0, \frac{1}{2}, 1, \frac{3}{2}, 2, \dots$ 
The value of $s$, together with the eigenvalue of a component of $M_{ik}$, say $M_{12}$ with eigenvalue $s_3$,
where $s_3 = -s, -s+1,\dots, s-1, s$, label a complete set of eigenstates of the angular momentum $\mathbf{M}$.
In single-particle representations, $\mathbf{M}$ corresponds to the spin of the particles.

Within the state space of an irreducible representation, a basis can be defined whose states are labelled by 
their eigenvalues with respect to a complete set of commuting operators; for example, 
the operators of the linear momentum $p_\mu$ and a component of $w_\mu$, say $w_3$.
Because the Casimir operator $P = p^\mu p_\mu$ has a constant value, only the three spatial components 
$\mathbf{p}$ of $p_\mu$ are independent.
Given that by Equation (\ref{1-8}) the eigenvalue of $W$ fixes the value $s$ of the angular momentum, we 
additionally need the eigenvalue $s_3$ of $M_{12}$ to label a complete set of basis states (in Dirac's bra-ket 
notation) $\left|\mathbf{p}, s_3 \right>$.

There are two irreducible representations for each value of $P$ and $W$, one for each sign of $p_0/|p_0|$,
which commutes with all the infinitesimal generators and is therefore an invariant of the group.

\subsection{Product representations and their reduction}

According to the axioms of quantum mechanics, a system of $N$ independent particles is described by the tensor 
product of single-particle representations (see, for instance, \cite{jmj}).
In such a product representation, the total linear momentum and the angular momentum can be defined by
\begin{equation}
p_\mu = \sum^N_{i=1} p^i_\mu  \;\;   \mbox{ and }    \;\;  M_{\mu \nu} = \sum^N_{i=1} M^i_{\mu \nu} , \label{2-1}
\end{equation}
where $p^i_\mu$ and $M^i_{\mu \nu}$ are operators of the representation of particle i.
It is easy to verify that the multi-particle operators (\ref{2-1}) satisfy the same
commutation relations (\ref{1-1}--\ref{1-3}) as the single-particle operators.
Therefore, a product representation has the same number of Casimir operators as a single-particle representation
and expressed by the operators (\ref{2-1}) they have the same form.

By construction, a product representation of the Poincar\'e group is not irreducible.
It can, however, be reduced to a direct sum of irreducible representations, as first described by Joos \cite{hj}. 
Reduction essentially means sorting states according to the eigenvalues of the Casimir operators. 
The different irreducible representations can then be classified by the eigenvalues of the Casimir operators 
$P$ and $W$ built from the multi-particle operators defined in (\ref{2-1}).
Therefore, a multi-particle system described by an irreducible representation has an invariant
effective mass $m_{\mathrm{tot}}$, with $m_{\mathrm{tot}}^2 = P$, and an invariant product of total linear 
momentum and angular momentum, which in the rest frame can be written as $m_{\mathrm{tot}}^2\,\mathbf{M}^2 = W$.
The angular momentum $\mathbf{M}$ now consists of the particle spins and the orbital angular momentum 
that corresponds to a rotation of the N-particle system as a whole.

In the following, for reasons of simplicity, I will ignore spin variables.
Therefore, the angular momentum will be identical to the orbital angular momentum and the individual 
particle states will be labelled solely by their linear momenta.

It should be emphasised that by Equation (\ref{1-8}) the orbital angular momentum is tightly anchored in 
the Casimir operator $W$, which means that it is an inherent property of an irreducible representation of the 
Poincar\'e group.
It is the multi-particle equivalent of the spin of an elementary particle.

As in the single-particle case, a basis of eigenstates $\left|\mathbf{p}, s_3 \right>$ of the (total) linear 
momentum and a component of the angular momentum can be chosen.
Due to the double function of the infinitesimal generators as observables and as generators of symmetry 
transformations, an eigenstate of the orbital angular momentum has a rotationally symmetric structure.
This means that in irreducible $N$-particle representations, the individual particle states are forced 
into a geometrically correlated configuration.
This is in sharp contrast to what would be expected from a classical point of view for a system of 
``independent'' particles.

\subsection{Two-particle representations}

In the following, I will study the consequences of this correlation for two-particle systems.

Consider two independent particles with momenta $\mathbf{p}_1$ and $\mathbf{p}_2$, prepared in a (tensor) product 
state $\left|\mathbf{p}_1\right>\!\left|\mathbf{p}_2 \right>$.
Following a common habit, I shall write this product state also in the form $\left|\mathbf{p}_1,\mathbf{p}_2 \right>$.
The product states are normalised via the inner product $\left<..|..\right>$ by
\begin{equation}
\left<\mathbf{p}_1,\mathbf{p}_2 | \mathbf{p}'_1,\mathbf{p}'_2\right> = 
\delta(\mathbf{p}_1 - \mathbf{p}'_1) \, \delta(\mathbf{p}_2 - \mathbf{p}'_2) .      							\label{2-2}
\end{equation}
By superimposing these product states, the basis of eigenstates $\left|\mathbf{p},s_3\right>$ of the total linear momentum 
and the angular momentum is formed.
Because of their rotational symmetry, the basis states are a superposition of product states, 
such that along with any pure product state $\left|\mathbf{p}_1,\mathbf{p}_2\right>$, the rotated versions of this 
state also contribute to the superposition.
 
This consideration largely determines the structure of the basis states. 
They are a momentum-entangled, rotationally symmetrical superposition of product states $\left|\mathbf{p}_1,\mathbf{p}_2\right>$
with the same total momentum $\mathbf{p} = \mathbf{p}_1 + \mathbf{p}_2$
\begin{equation}
\left|\mathbf{p},s_3\right>\; = \;\int_\Omega\!d^3\mathbf{p}_1 d^3\mathbf{p}_2\; 
\delta(\mathbf{p} - \mathbf{p}_1 - \mathbf{p}_2)\;
c_{\mathbf{p}, s_3}(\mathbf{p}_1,\mathbf{p}_2) \, 
\left|\mathbf{p}_1,\mathbf{p}_2\right> .                                           								\label{2-4}
\end{equation} 
The integration area $\Omega$ is part of the two-particle mass shell
\begin{equation}
(p_1 + p_2)^\mu (p_1 + p_2)_\mu  = m_{\mathrm{tot}}^2 .                                              			\label{2-5}
\end{equation} 

With the alternative variable set $\mathbf{k}$ = $\mathbf{p}_1 - \mathbf{p}_2$ and $\mathbf{p}$ = $\mathbf{p}_1 + \mathbf{p}_2$  
we have $\mathbf{p}_1 = \frac{1}{2}(\mathbf{p+k})$ and $\mathbf{p}_2 = \frac{1}{2}(\mathbf{p-k})$; 
therefore the state (\ref{2-4}) can be given in the form
\begin{equation}
\left|\mathbf{p},s_3\right>\; = \;\int_\Omega\!d^3\mathbf{k}\; 
\left|\frac{\mathbf{p+k}}{2}\right> \,
c_{\mathbf{p}, s_3}(\mathbf{k}) \, 
\left|\frac{\mathbf{p-k}}{2}\right> ,                                               						 \label{2-6}
\end{equation}  
where I have written the product state $\left|\mathbf{p}_1,\mathbf{p}_2\right>$ explicitly as a tensor product.

This form suggests a familiar physical interpretation:
The state describes two particles {\it coupled by the field} $c_{\mathbf{p}, s_3}(\mathbf{k})$, which controls
(causes) a {\it virtual exchange of momentum} $\mathbf{k}$ between the particles.  
This is formally the same as the description of the interaction mechanism of the Standard Model, where a similar 
field is understood as a gauge field and the exchanged momenta as virtual gauge bosons.
Because all basis states have the same form, ``virtual exchange of momentum'' is a general structural element of 
irreducible two-particle representations of the Poincar\'e group.

\subsection{Normalisation of two-particle states}

The two-particle state (\ref{2-4}) still needs to be normalised according to the standard rule of quantum mechanics: 
If $N$ normalised orthogonal states are superimposed, then the resulting state must be re-normalised with the factor 
$N^{-\frac{1}{2}}$.
If the summation is replaced by an integration, then the normalisation factor is $\omega = V\!(\Omega)^{-\frac{1}{2}}$, 
where $V$ is the volume of the region of integration $\Omega$.

A general state (i.e. not necessarily an eigenstate) of the irreducible two-particle representation,
normalised by the factor $\omega$, can then be written in the form
\begin{equation}
\left|\phi\right>\; = \;\int_\Omega\!\omega\,d^3\mathbf{p}_1 d^3\mathbf{p}_2\; 
c_\phi(\mathbf{p}_1,\mathbf{p}_2) \, \left|\mathbf{p}_1,\mathbf{p}_2\right>.         	 						\label{2-8}
\end{equation} 
With the normalisation condition of product states (\ref{2-2}), the inner product evaluates to
\begin{eqnarray}
\left<\phi|\phi\right> &=& \int_\Omega\!\omega^2\,d^3\mathbf{p}_1 d^3\mathbf{p}_2 d^3\mathbf{p'}_1 d^3\mathbf{p'}_2\;
c^*_\phi(\mathbf{p}_1,\mathbf{p}_2) c_\phi(\mathbf{p'}_1,\mathbf{p'}_2) 
\left<\mathbf{p}_1,\mathbf{p}_2 | \mathbf{p}'_1,\mathbf{p}'_2\right> \;\;\;\; \label{2-10} \\
&=& \int_\Omega\!\omega^2\,d^3\mathbf{p}_1 d^3\mathbf{p}_2\;c^*_\phi(\mathbf{p}_1,\mathbf{p}_2) c_\phi(\mathbf{p}_1,\mathbf{p}_2)
\;\;!= 1.\label{2-11}
\end{eqnarray}
The last equation (\ref{2-11}) gives the normalisation condition for the coefficients $c_\phi(\mathbf{p}_1,\mathbf{p}_2)$.

I will call the infinitesimal volume element $\omega\,d^3\mathbf{p}_1 d^3\mathbf{p}_2$ of the integral (\ref{2-8}) 
the {\it normalised volume element} on $\Omega$.
For reasons of transparency, I will extract the normalisation factor $\omega$ from the state and write a normalised 
two-particle state in the following as $\omega \left|\phi\right>$.

In Section 4, I will calculate the numerical value of the normalisation factor $\omega = V(\Omega)^{-\frac{1}{2}}$.
This calculation is anything but trivial due to the specific topology and the non-Euclidean metric of the 
two-particle mass shell.

\section{A gedanken experiment}
\label{STAN}

The following scattering experiment illustrates the physical effects of the virtual momentum exchange described by
Equation (\ref{2-6}).

Suppose two independently prepared particles, described by a product state $\left|\mathbf{p}_1,\mathbf{p}_2\right>$, 
pass through a measuring device capable of measuring their orbital angular momentum while leaving the total momentum unchanged. 
The device will measure an angular momentum $s$, and leave the two-particle system in the 
state $\omega \left|\mathbf{p},s_3\right>$.
This state is then analysed by independently measuring the outgoing particle momenta $\mathbf{p}'_1$ and $\mathbf{p}'_2$.
The total transition amplitude between an incoming product state $\left|\mathbf{p}_1,\mathbf{p}_2\right>$
and an outgoing product state $\left|\mathbf{p}'_1,\mathbf{p}'_2\right>$ is given by 
\begin{equation}
S = \omega^2 \left<\mathbf{p}'_1, \mathbf{p}'_2|\mathbf{p},s_3\right>\!
\left<\mathbf{p},s_3|\mathbf{p}_1,\mathbf{p}_2\right> .                                    					\label{3-1}
\end{equation} 
Since in the intermediate state the information about the incoming particle momenta is lost, the outgoing particle 
momenta will, in general, be different from the incoming momenta.
In other words, we will observe a scattering with an amplitude determined by $\omega^2$, 
which here acts like a coupling constant.

This scattering mechanism can be seen as the two-particle analogue of diffraction of a plane wave at a pinhole.
While a pinhole selects a spherical elementary wave, this gedanken experiment selects an eigenstate of the orbital 
angular momentum.

\section{Calculating the normalisation factor}
\label{CALC}

The construction of the normalised infinitesimal volume element on the two-particle mass shell was first described in \cite{sm3}. 
This construction involved rather abstract geometrical considerations on the basis of the Lie ball in 5 dimensions. 
It remained, therefore, non-transparent with respect to its physical contents and consequently not really understood.
The present paper provides a simpler, more detailed, and, above all, physically transparent description of that construction.

\subsection{The two-particle mass shell $\mathcal{R}_{shell}$}
\label{MASSSH}

Let $p_1$ and $p_2$ be the 4-momenta of two particles with masses $m_1$ and $m_2$.
They satisfy the mass-shell relations
\begin{equation}
{p_1}^2 = m_1^2 \;\; \mbox{ and } \;\; {p_2}^2 = m_2^2 .                          								\label{4-1}
\end{equation}               
The total momentum $p$ and the relative momentum $q$ are defined by
\begin{equation}
p = p_1 + p_2 \;\; \mbox{ and } \;\; q = p_1 - p_2                                								\label{4-2}
\end{equation}
and satisfy
\begin{equation}
p\,q = m_1^2 - m_2^2 .                                                            								\label{4-3}
\end{equation}
Equation (\ref{4-3}) allows rotating the vector $q$ relatively to $p$ by the action of $SO(3)$, as long as no other 
restrictions apply.

For an irreducible two-particle representation,
\begin{equation}
p^2 = m_{\mathrm{tot}}^2                                                                   				\label{4-4}   
\end{equation}
and
\begin{equation}
q^2 = 2 m_1^2 + 2 m_2^2 - m_{\mathrm{tot}}^2,                                       							\label{4-5}
\end{equation}
where $m_{\mathrm{tot}}^2$ is the constant value of the Casimir operator $P$.

Unless a particle is combined with an anti-particle, $q$ is space-like. 
Then $q^2 < 0$ and by Equation (\ref{4-5}) 
\begin{equation}
m_{\mathrm{tot}}^2 > 2 m_1^2 + 2 m_2^2, 																													\label{4-8}
\end{equation}
corresponding to scattering states. 
If a particle is combined with an anti-particle, i.e. with $p_2$ on the negative half of its mass shell, 
$q$ is time-like. 
Then $q^2 > 0$ and 
\begin{equation}
m_{\mathrm{tot}}^2 < 2 m_1^2 + 2 m_2^2, 																													\label{4-9}
\end{equation}
indicating the possibility of bound states.

Equations (\ref{4-4}) and (\ref{4-5}) can be combined to form the equation of the two-particle mass shell 
\begin{equation}
p^2 + q^2 =  2 m_1^2 + 2 m_2^2 .                                                  								\label{4-6}   
\end{equation}
The symmetries of this equation require special attention.
Rotations around the 3-momentum $\mathbf{p}$ as the axis leave $p$ invariant, but change $q$. 
Therefore, these rotations define an internal degree of freedom with an $SO(2)$ symmetry.
The $SO(2)$ moves within $SO(3,1)$ as $p$ moves through the hyperboloid (\ref{4-4}).
The two-particle mass shell has, therefore, the structure of a circle bundle over the mass hyperboloid of $p$.
The circle fibres parametrise the internal rotational degree of freedom of a two-particle state.
Together, $SO(3,1)$ and the local $SO(2)$ generate the irreducible parameter space 
\begin{equation}
\mathcal{R}_{shell} = \{\mathbf{p} \in \mathbb{R}^3, \mathbf{q} \in \mathbb{R}^2; \,
\mathbf{p}^2 + \mathbf{q}^2 = p_0^2 + q_0^2 - 2 m_1^2 - 2 m_2^2 \} \;             								\label{4-7}   
\end{equation}
defined in terms of the independent parameters of the two particles.

The local $SO(2)$ degree of freedom is in contrast to the $SO(3)$ degree of freedom of $q$ in Equation (\ref{4-3}), 
which takes into account only the constancy of the Casimir operator $P$, but disregards the second 
Casimir operator $W$.

In the following section, I will examine how the geometry of $\mathcal{R}_{shell}$ is related to standard 
geometrical elements, such as spheres and balls.

\subsection{The homogeneous domain $\mathcal{R}_{\mbox{\tiny{IV}}}$}
\label{HOMO}

The space $\mathcal{R}_{shell}$ has an $SO(3,1)$ symmetry under rotations and boost operations of $p$, 
an $SO(2,1)$ symmetry under rotations and boost operations of $q$,
and a merely apparent $SO(2)$ symmetry under rotations in the $p_0$--$q_0$ plane.
If we could ``rotate'' (as explained below) a component of $\mathbf{p}$ into a component of $\mathbf{q}$, 
then the equation would have a full $SO(5,2)$ symmetry.
As it is, the physically relevant symmetries are the $SO(3,1)$ symmetry and the local $SO(2)$ symmetry. 

The fact that the group $SO(5,2)$ differs from their combined subgroups $SO(3,1), SO(2,1)$, and $SO(2)$ by only one 
additional rotation group $SO(2)$ suggests that the volume of the infinitesimal volume element on $\mathcal{R}_{shell}$ 
can be derived from the corresponding volume element on a symmetric space for the full $SO(5,2)$.

Such spaces, more precisely, bounded homogeneous domains, have been studied by Cartan \cite{eca}. 
Cartan has proved that there exist 6 types of bounded homogeneous domains.
One of these domains, $\mathcal{R}_{\mbox{\tiny{IV}}}$, is suited for our purposes.

The domain $\mathcal{R}_{\mbox{\tiny{IV}}}$ of $n$-dimensional complex vectors $z = x + iy$ can be realised 
on the Lie ball (cf. Hua \cite{hua1})
\begin{equation}
D^n = \{z \in \mathbb{C}^n; 1 + |zz'|^2-2\bar{z}z' > 0, |zz'| < 1 \} .              							\label{5-1}
\end{equation}
Its boundary (the Shilov boundary) is given by
\begin{equation}
Q^n = \{\xi = x\,e^{i\theta}; x \in \mathbb{R}^n, xx'=1 \}, \;
0<\theta<\pi .                                                                      							\label{5-2}
\end{equation}
The vector $z'$ is the transpose of $z$, $\bar{z}$ is the complex conjugate of $z$.

Unfortunately, from the physicist's point of view, the realisation of $\mathcal{R}_{\mbox{\tiny{IV}}}$ on the Lie ball 
lacks transparency.
However, as Hua \cite{hua2} has shown, $\mathcal{R}_{\mbox{\tiny{IV}}}$ can also be regarded as a homogeneous space of 
$2 \times n$ real matrices  
\begin{equation}
A = \left| \begin{array}{lllll} x_1 & x_2 & x_3 & ... & x_n  \vspace{0.2cm} \\ 
                                y_1 & y_2 & y_3 & ... & y_n  \end{array} \right |\;                       
\end{equation}
built from the real and imaginary parts of $z$. 
In matrix form, the homogeneous domain $\mathcal{R}_{\mbox{\tiny{IV}}}$ is defined by 
\begin{equation} 
D^n = I^{(2)} - A A' > 0 ,                                                          							\label{5-3}
\end{equation}
where $I^{(2)}$ is the unit matrix in two dimensions and $A'$ the transposed of $A$.
The boundary of $D^n$ is 
\begin{equation} 
Q^n = I^{(2)} - A A' = 0 .                                                          							\label{5-4}
\end{equation}

The realisation of the symmetry group $SO(5,2)$ on the bounded $D^5$ involves a Cayley transformation, which maps the 
unbounded $\mathbb{R}^5\times\mathbb{R}^5$ onto $D^5$.
In physical applications, the transformation of the infinitely extended momentum space onto a bounded domain is more 
confusing than helpful. 
Therefore, I will use the natural realisation of $SO(5,2)$ on the unbounded $\mathbb{R}^5\times\mathbb{R}^5$ with
$D^5$ as the associated unit ball.

To establish the connection with $\mathcal{R}_{shell}$, consider a 3-dimensional vector $\mathbf{x}$, 
proportional to $\mathbf{p}$, and a 2-dimensional vector $\mathbf{y}$, proportional to $\mathbf{q}$, both rewritten as 
vectors in $\mathbb{R}^5$,
\begin{equation}
\mathbf{x} = (x_1,\,x_2,\,x_3,\;0,\;0) \; \mbox{ and } \; \mathbf{y} = (0,\;0,\;0,\,y_4,\,y_5).		\label{5-5}
\end{equation}
These vectors can be combined into the $2 \times 5$ matrix 
\begin{equation} 
A = \left| \begin{array}{lllll} x_1 & x_2 & x_3 & \,0 & \,0  \vspace{0.2cm} \\ 
                                \,0 & \,0 & \,0 & y_4 & y_5  \end{array} \right | .              	\label{5-6} 
\end{equation}
The matrix product $A A'$ can be evaluated as 
\begin{equation}
A A' = \left| \begin{array}{ll} \mathbf{x}^2 & \,0 \vspace{0.2cm} \\ \,0 & \mathbf{y}^2 \end{array} \right|.  \label{5-7}
\end{equation}
Hence, the unit ball $D^5$ can be rewritten as the direct product of two unit balls
\begin{equation}
\mathbf{x}^2 < 1  \;\; \mbox{ and } \;\; \mathbf{y}^2 < 1  																				\label{5-8}
\end{equation}
on $\mathcal{R}_{shell}$.

The meaning of a ``rotation'' of a component of $\mathbf{p}$ into a component of $\mathbf{q}$ is now obvious: if it were 
possible to rotate the component $x_3$ into $x_4$ and $y_4$ into $y_3$ by the operations of a single additional $SO(2)$ group, 
then both vectors, $\mathbf{x}$ and $\mathbf{y}$, would be converted into true 5-dimensional vectors and, together with the  
existing $SO(3)$ symmetry of $\mathbf{p}^2$ and the $SO(2)$ symmetry of $\mathbf{q}^2$, the whole $D^5$ would be covered.

The Euclidean volumes of the unit ball $D^5$ and the unit sphere $Q^5$ have been calculated in Hua \cite{hua1}, with the 
results 
\begin{equation}
V\!(D^5) = \frac{\pi^5}{2^4\, 5!}\;,\;\;\; V\!(Q^5) = \frac{8 \pi^3}{3} .                         \label{5-9}
\end{equation}

\subsection{The normalised volume element on $\mathcal{R}_{shell}$}
\label{NORMA}

Now we are prepared to answer the question: Given a normalised spherical volume element on $\mathcal{R}_{\mbox{\tiny{IV}}}$, 
what will be the corresponding Euclidean volume element on $\mathcal{R}_{shell}$?

The volume of the unit ball $D^5$ can be expressed by an integral over $D^5$.
This integral can be split into an integral over the unit sphere $Q^5$ 
and a second integral over the radial direction of $D^5$:
\begin{equation}
V\!(D^5) = \int_0^1 dr\;r^4\!\int_{Q^5} d^4x .                                                  	\label{6-2}                       
\end{equation}
The infinitesimal volume element $d^4x$ in the integral over $Q^5$, resulting in the volume $V\!(Q^5)$, is normalised 
by the inverse of the square root of this volume.

The unit sphere $Q^5$ can be understood as generated by the action of $SO(5)$, whereas the corresponding area on 
$\mathcal{R}_{shell}$ can be understood as generated by the combined actions of $SO(3)$ and a local $SO(2)$, equivalent to 
the actions of four rotations with orthogonal axes, that is, the group $SO(4)$. 
Hence, compared to the volume element $d^4x$ on $Q^5$, the corresponding volume element on $\mathcal{R}_{shell}$ not only 
has a different size, but also a dimension that has been reduced by one.
The ratio of these two volumes is given by the volume of the symmetric space $SO(5)/SO(4)$, which is the unit sphere in five 
dimensions $S^4$.
Dividing $d^4x$ by $V\!(S^4)$ reduces the volume calculated by the integral over $D^5$ to the corresponding volume on 
$\mathcal{R}_{shell}$.

The infinitesimal volume element $dr\,d^4x$ is still a spherical volume element with equal scalings in four directions, 
but with a different scaling in the radial direction.
To give the infinitesimal volume element the form of an isotropic Cartesian volume element, 
as is used in the integral of the two-particle state (\ref{2-8}), we first have to replace $D^5$ 
by a cuboid with the same volume $V\!(D^5)$ and an edge length of 1 in a formerly radial direction.
On this cuboid, the integration over $r$ delivers a factor of 1 to the volume of the cuboid.
Together with the integration over the four other directions, the infinitesimal volume elements add up to the volume $V\!(D^5)$
of the original Lie ball.
Therefore, each of the four volume elements $dx_1, dx_2, dx_3, dx_4$, now considered as part of a Cartesian volume element, 
contributes a factor of $V\!(D^5)^{\frac{1}{4}}$ to this volume.
To make the five-dimensional volume element an isotropic one, $dr$ must be rescaled by the factor $V\!(D^5)^{\frac{1}{4}}$.

Finally, the Jacobian that relates $p$ and $q$ to $p_1$ and $p_2$ (cf. Equation (\ref{4-2})) contributes an additional factor of 2.

Taking all factors together results in the volume term
\begin{equation}
\omega^2 = 2\,V\!(D^5)^{\frac{1}{4}} \, / \, (V\!(Q^5)\,V\!(S^4)) .                      					\label{6-3} 
\end{equation}
Its square root, the factor $\omega$, normalises the Euclidean infinitesimal volume element on $\mathcal{R}_{shell}$. 
It is determined, at first, for a single point on the unit sphere $Q^5$, but since $\mathcal{R}_{\mbox{\tiny{IV}}}$ is a 
homogeneous symmetric domain, the same factor is obtained everywhere on $\mathcal{R}_{\mbox{\tiny{IV}}}$. 
Hence, $\omega$ is a constant: its numerical value is characteristic for an irreducible two-particle representation of the 
Poincar\'e group.

I have to come back to the gedanken experiment resulting in the transition amplitude (\ref{3-1}). 
In this scattering experiment, only the value of the first Casimir operator $P$ is determined by the incoming product state.
Therefore, all intermediate states $\left|\mathbf{p},s_3\right>$ with the same values of $\mathbf{p}$ and $(s, s_3)$, but different 
values of the second Casimir operator $W$, may contribute to the transition amplitude. 
Their amount derives from comparing the range of $q$ within a product representation, as covered by the action of $SO(3)$ 
(cf.\,Equation (\ref{4-3})), 
with its range within an irreducible representation, as covered by the action of $SO(2)$.
This ratio is given by $4\pi$, which is the volume of the symmetric space $SO(3)/SO(2)$, which is equal to the unit sphere in three 
dimensions $S^2$.
If we take into account all irreducible representations, the constant $\omega^2$ in the transition amplitude (\ref{3-1}) must 
be adjusted by this ratio, leading to the actual coupling constant
\begin{equation}
4\pi\,\omega^2 = 8\pi\,V\!(D^5)^{\frac{1}{4}} \, / \, (V\!(Q^5)\,V\!(S^4)) .             					\label{6-4}
\end{equation}

Inserting the explicit values (taken from Hua \cite{hua1})
\begin{equation}
V\!(D^5) = \frac{\pi^5}{2^4\, 5!},\;\;    	 
V\!(Q^5) = \frac{8 \pi^3}{3},\;\;         	 
V\!(S^4) = \frac{8 \pi^2}{3}                                               	   
\end{equation}
leads to 
\begin{equation}
4\pi\,\omega^2 \;= \; \frac{9}{16 \pi^3} \left(\frac{\pi}{120}\right)^{1/4}              					\label{6-5}
= \; 1/137.03608245 ,   			          	 
\vspace{0.2cm}
\end{equation}
which closely matches the CODATA value $1/137.035999139$ \cite{cod} of the electromagnetic fine-structure constant $\alpha$.
Note that the theoretical value (\ref{6-5}) refers to a universe with only two particles; it certainly does not include 
``hadronic corrections''.

This agreement establishes a connection between the fine-structure constant and the normalisation constant of the two-particle 
states by
\begin{equation}
\alpha \simeq 4\pi\,\omega^2 .   	                                                       					\label{6-6}
\vspace{0.2cm}
\end{equation}
The gedanken experiment in Section 3, which identifies $\omega^2$ as a coupling constant, shows that this relation is not 
just a numerical coincidence.

The representation of $\alpha$ by the term $8\pi\,V\!(D^5)^{\frac{1}{4}} \, / \, (V\!(Q^5)\,V\!(S^4))$ was accidentally
found in 1971 by the Swiss mathematician Armand Wyler (a former student of Heinz Hopf), while he was examining some
symmetric domains. 
It has become known as ``Wyler's formula'' for the fine-structure constant.
Wyler \cite{aw} published his finding, hoping to attract the interest of physicists. 
Unfortunately, Wyler was not able to put his observation into a convincing physical context. 
Therefore, the physics community dismissed his formula as meaningless numerology (see, for example, \cite{meh}). 
Now it has become evident that Wyler's formula directly links the signature of the electromagnetic interaction 
with the geometric footprint of the fibred two-particle mass shell.

\section{Conclusions}
\label{CONC}

The simple and seemingly insignificant question about the normalisation of two-particle states of irreducible 
two-particle representations of the Poincar\'e group has led to the relation $\alpha = 4\pi\, \omega^2$ between 
the normalisation factor $\omega$ and the electromagnetic fine structure constant $\alpha$.
Within these states, the same factor---which is now considered to be a coupling constant---determines the strength of 
an interaction between the particles, caused by the constraints of an irreducible representation.
The matching values of the coupling constants strongly suggest that this interaction and the phenomenological 
electromagnetic interactions are one and the same. 
The mathematical similarity of the basic interaction mechanisms---here ``interaction by exchange of momentum'', 
there ``interaction by exchange of gauge bosons''---indicates that in the classical limit the ``exchange of momentum''  
will also lead to the Maxwell equations, especially since the structure of the Maxwell equations is largely 
determined by the Lorentz symmetry.

This would mean that on the quantum mechanical level, the electromagnetic interaction is a geometry-induced interaction  
that is based solely on the structure of the Poincar\'e group.

These insights into the geometry of two-particle states should have far-reaching consequences for the interpretation 
of the Standard Model and, in particular, for its extension by a quantum theory of gravity.

\section*{Acknowledgements}
\label{ACKN}

This work would not have been possible without Wyler's discovery.
Without his formula, I would never have embedded the two-particle mass shell into one of Cartan's homogeneous domains.
Without this embedding, I would not have found the link between the geometry of the two-particle mass shell and $\alpha$.
And without the connection to $\alpha$ it would never have come to my mind that two ``independent'' particles ``interact'' 
with each other when configured as a two-particle system with well-defined total linear and angular momentum.

\end{document}